\documentclass[prl,showpacs,twocolumn]{revtex4}%
\input{psfig.sty}
\usepackage{amsmath}
\usepackage{graphicx}
\usepackage{amsfonts}
\usepackage{amssymb}%

\begin{document}
\title{Observation of dark state polariton collapses and revivals}
\date{\today }
\author{ D. N. Matsukevich, T. Chaneli\`{e}re, S. D. Jenkins, S.-Y. Lan, T.A.B. Kennedy, and A. Kuzmich}
\affiliation{School of Physics, Georgia Institute of Technology, Atlanta, Georgia 30332-0430}\pacs{03.65.Ud,03.67.Mn,42.50.Dv}
\begin{abstract}
By time-dependent variation of a control field, both coherent and single photon states of light are stored in, and retrieved from, a cold atomic
gas. The efficiency of retrieval is studied as a function of the storage time in an applied magnetic field. A series of collapses and revivals is
observed, in very good agreement with theoretical predictions. The observations are interpreted in terms of the time evolution of the collective
excitation of atomic spin wave and light wave, known as the dark-state polariton.
\end{abstract}
\maketitle

Atomic ensembles show significant promise as quantum memory elements in a quantum network \cite{briegel,knill,duan,kuzmich,lukin}. A ``dark-state
polariton" is a bosonic-like collective excitation of a signal light field and an atomic spin wave \cite{fleischhauer}, whose relative amplitude
is governed by a control laser field. In the context of quantum memories, the dark state polariton should enable adiabatic transfer of single
quanta between an atomic ensemble and the light field. Seminal ``stopped-light" experiments that used laser light excitation
\cite{phillips,hau,mair} can be understood in terms of the dark-state polariton concept. In a recent work the storage and retrieval of single
photons using an atomic ensemble based quantum memory was reported, and the storage time was conjectured to be limited by inhomogeneous
broadening in a residual magnetic field \cite{chaneliere}.

We have recently predicted that dark-state polaritons will undergo collapses and revivals in a uniform dc magnetic field \cite{jenkins}. During
storage, the dark-state polariton consists entirely of the collective spin wave excitation. According to the dark-state polariton concept, the
retrieved signal field should exhibit the collapse and revivals experienced by the spin wave. The revivals occur at integer multiples of one half
the Larmor period, with dynamics that are sensitive to the relative orientation of the magnetic field and the light wavevector. The spin wave
part of the dark-state polariton involves a particular superposition of atomic hyperfine coherences (see Eq.(1) below), intimately related to the
phenomenon of electromagnetically-induced transparency (EIT) \cite{harris,scully}. Revivals of single atom coherences were observed in atom
interferometery \cite{pritchard,smith}. Coupled exciton-polariton oscillations in semiconductor microcavities have also been reported
\cite{weisbuch,jacobson}.

The remarkable protocol of Duan, Lukin, Cirac, and Zoller (DLCZ) provides a measurement-based scheme for the creation of atomic spin excitations
\cite{duan}. In systems where EIT is operative, these excitations will in general contain a dark-state polariton component. The orthogonal
contribution may be regarded as a bright-state polariton in that it couples dissipatively to the excited atomic level in the presence of the
control field \cite{fleischhauer1}. Observation of the retrieved signal field, however, picks out the dark state polariton part, while the
orthogonal component is converted into spontaneous emission \cite{jenkins}. A number of previous works reported generation and subsequent
retrieval of DLCZ collective excitations \cite{kuzmich2,eisaman,jiang,eisaman1,chou,matsukevich,balic,matsukevich1,felinto}. Several of these
studies investigated the decoherence of these excitations in cold atomic samples \cite{kuzmich2,chou,matsukevich,matsukevich1,felinto}. It has
been similarly conjectured in these works that the decay of the coherence was due to spin precession in the ambient magnetic field.  While the
observed decoherence times are consistent with the residual magnetic fields believed to be present, the observation of revivals predicted in
Ref.\cite{jenkins} would be solid proof that Larmor precession is indeed the current limitation on the quantum memory lifetime. Moreover,
controlled revivals could provide a valuable tool for quantum network architectures that involve collective atomic memories
\cite{duan,kuzmich,lukin}.
\begin{figure}[htp]
\begin{center}
\leavevmode  \psfig{file=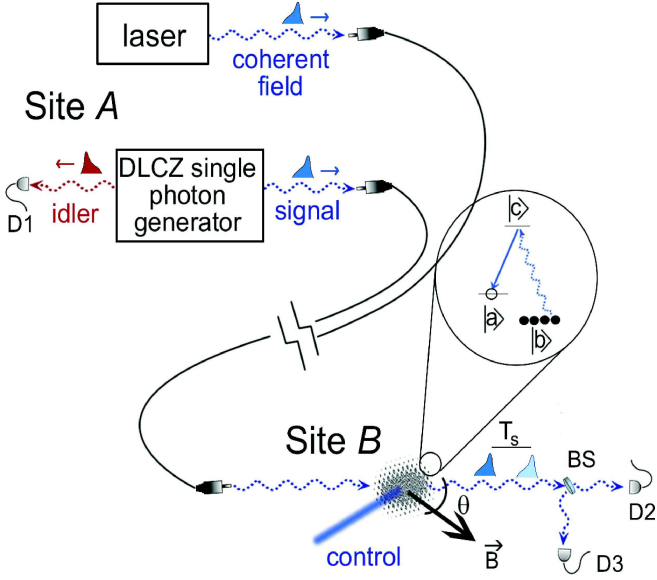,height=3.0in,width=3.4in}
\end{center}
\caption{ A schematic diagram illustrates our experimental setup. A signal field from either a laser, or a DLCZ source of conditional single
photons at Site {\it A} is carried by a single-mode fiber to an atomic ensemble at Site {\it B}, where it is resonant on the $|b\rangle
\leftrightarrow |c\rangle$ transition. The signal field is stored, for a duration $T_s$, and subsequently retrieved by time-dependent variation
of a control field resonant between levels $|a\rangle$ and $|c\rangle$. All the light fields responsible for trapping and cooling, as well as the
quadrupole magnetic field in the MOT, are shut off during the period of the storage and retrieval process. An externally applied magnetic field
created by three pairs of Helmholtz coils (not shown) makes an angle $\theta $ with the signal field wavevector. The inset shows the structure
and the initial populations of atomic levels involved. The signal field is measured by detectors D2 and D3, while detector D1 is used in the
conditional preparation of single photon states of the signal field at Site {\it A}.}\label{F1}
\end{figure}

With this goal in mind, we report in this Letter observations of collapses and revivals of dark-state polaritons in agreement with the
theoretical predictions \cite{jenkins}. In our experiment, we employ two different sources for the signal field, a coherent laser output and a
conditional source of single photons \cite{duan}. The latter is achieved by using a cold atomic cloud of $^{85}$Rb at Site {\it A} in the
off-axis geometry pioneered by Harris and coworkers \cite{balic}.  Another cold atomic cloud of $^{85}$Rb at Site {\it B} serves as the atomic
quantum memory element, as shown in Fig.~1. Sites {\it A} and {\it B} are physically located in adjacent laboratories connected by a 100 meter
long single-mode optical fiber. The fiber channel directs the signal field to the optically thick atomic ensemble prepared in level $|b\rangle $.
The inset in Fig.~1 indicates schematically the structure of the three atomic levels involved, $|a\rangle ,|b\rangle $ and $|c\rangle $, where
$\{|a\rangle;|b\rangle \}$ correspond to the $5S_{1/2},F_a=3, F_b=2$ levels of $^{85}$Rb, and $|c\rangle$ represents the $5P_{1/2},F_c=3$ level
associated with the $D_1$ line at 795 nm. The signal field is resonant with the $|b\rangle \leftrightarrow |c\rangle $ transition and the control
field with the $|a\rangle \leftrightarrow |c\rangle $ transition.

When the signal field enters the atomic ensemble at Site {\it B}, its group velocity is strongly modified by the control field. By switching off
the control field within about 100 ns, the coupled excitation is converted into a spin wave excitation with a dominant dark state polariton
component, i.e., the signal field is ``stored" \cite{fleischhauer,phillips,hau,mair}. An important condition to achieve this storage is a
sufficiently large optical thickness of the atomic sample, which enables strong spatial compression of the incident signal field \cite{lukin}. In
our experiment the measured optical thickness $d\simeq 8$. The subsequent evolution of a dark state polariton in an external magnetic field is
predicted to reveal a series of collapses and revivals whose structure is sensitive to the magnitude and orientation $\theta $ of the field
relative to the signal wavevector \cite{jenkins}.

As we deal with an unpolarized atomic ensemble, we must take into account the Zeeman degeneracy of the atomic levels. Choosing the same circular
polarizations for both the probe and the control fields allows us to retain transparency \cite{chaneliere}. For a $\sigma_+$ polarized signal
field, the dark state polariton annihilation operator for wavenumber $q$ is given by \cite{jenkins}
\begin{eqnarray}
  \hat{\Psi}\left(q,t\right)
  = \frac{\Omega(t)\hat{a}_{k,+}-\sqrt{Np}g^{\ast}
          \sum_{m} R_{m}
           \hat{S}_{a~m}^{b~m}\left(  q,t\right)  }
    {\sqrt{\left\vert \Omega(t)\right\vert^{2} + Np \left\vert g\right\vert ^{2}
    \sum_{m}
       \left\vert R_{m}\right\vert ^{2}}}%
\end{eqnarray}
where $\Omega(t)$ is the control field Rabi frequency, $g$ the coupling coefficient for the signal transition, $m$ is the magnetic quantum
number, $R_m = C_{m~1~m+1}^{F_b~1~F_c} / C_{m~1~m+1}^{F_a~1~F_c}$ is a ratio of Clebsch-Gordan coefficients, $N$ is the number of atoms,
$p=1/(2F_b+1)$, $\hat{a}_{k,+}$ is the field annihilation operator for the mode of wavevector $k=q+\omega_0/c$ and positive helicity, $\omega_0$
is the Bohr frequency of the $|b \rangle \leftrightarrow |c\rangle $ transition, $S_{a~m}^{b~m}(q,t) \equiv 1/\sqrt{Np} \sum_{\mu}
\hat{\sigma}_{b\,m,~a\,m}^{(\mu)}(t) \exp(-i(qz_{\mu}-\Delta(t-z_{\mu}/c)))$ is a collective spin wave operator, where
$\hat{\sigma}_{b\,m,~a\,m}^{(\mu)}(0) = \vert b,m\rangle_{\mu}\langle a,m\vert$ is a hyperfine coherence operator for atom $\mu=1,..N$, $z_{\mu}$
is the position of atom $\mu$, and $\Delta$ is the hyperfine splitting of the ground state. When $R_m$ is finite for all $m$, the atomic
configuration supports EIT, but when one or more $R_m$ is infinite, there is an unconnected lambda configuration, EIT is not possible and dark
state polaritons do not exist. Specifically, the excited state $\vert c,m+1 \rangle$ is not coupled by the control field to a state in the ground
level $|a\rangle$. An atom in the state $\vert b,m \rangle$ would absorb the signal field as if no control field were present.

The signal is stored in the form of spin wave excitations associated with the dark state polaritons $\sim \sum_m R_m\hat{S}_{a~m}^{b~m}(q)$ for
some range of $q$'s. During the storage phase, and in the presence of the magnetic field $\mathbf{B}$, the atomic hyperfine coherences rotate
according to the transformation
\begin{eqnarray}
\hat{S}^{bm}_{am}(q,t) = \sum_{m_1=-F_b}^{F_b} \sum_{m_2=-F_{a}}^{F_{a}}\mathcal{D}^{(b)\dag}_{m_1 m} (t) \mathcal{D}^{(a)}_{m,m_2}
(t)\hat{S}^{bm_1}_{am_2}(q,0),
\end{eqnarray}
where $\mathcal{D}_{m,m^{\prime}}^{(s)}(t) \equiv \langle s,m\vert \exp(-ig_s\mathbf{\Omega}\cdot\mathbf{\hat{F}}t) \vert s,m^{\prime}\rangle$ is
the rotation matrix element for hyperfine level $s$, $\mathbf{\hat{F}}$ is the total angular momentum operator, $\mathbf{\Omega} \equiv
\mu_B\mathbf{B}/\hbar$, $\mu_B$ is the Bohr magneton, $g_a$ and $g_b$ are the Land\'{e} $g$ factors for levels $|a\rangle $ and $|b\rangle $ of
$^{85}$Rb and, ignoring the nuclear magnetic moment, $g_a = -g_b$. This rotation dynamically changes the dark state polariton population during
storage.

The measured signal retrieved after a given storage time $T_s$ is determined by the remaining dark state polariton population. Stated
differently, only the linear combination of hyperfine coherences $\sim \sum_m R_m\hat{S}_{a~m}^{b~m}(q,T_s)$ contributes to the retrieved signal.
We calculate the number of dark state polariton excitations as a function of $T_s$ using Eqs.(1) and (2), $\langle \hat{N}(T_s) \rangle = \sum_q
\langle\hat{\Psi}^{\dag}(q, T_s)\hat{\Psi}(q, T_s)\rangle$, and find
\begin{eqnarray}
  \frac{\langle \hat{N}(T_s)  \rangle}
       {\langle \hat{N}(0) \rangle}
   = \Bigg\vert \sum_{m_1 m_2}\frac{R_{m_1} R_{m_2}}
       {\sum_m
       \left\vert R_{m}\right\vert ^{2}}
  \mathcal{D}_{m_2 m_1}^{(b)  \dag}( T_s)
         \mathcal{D}_{m_1 m_2}^{(a)  }
                    \left( T_s\right)\Bigg\vert^2.
\end{eqnarray}

In Fig.~2, panels (f) through (j), we show the retrieval efficiency for various orientations of a magnetic field of magnitude 0.47 G,
corresponding to the Larmor period of 4.6 $\mu$s. With the approximation $g_a = -g_b$ it is clear that the system undergoes a revival to the
initial state after one Larmor period ($2\pi/\vert g_b\mathbf\Omega\vert$), and thus the signal retrieval efficiency equals the zero storage time
value. Depending on the orientation of the magnetic field, a partial revival at half the Larmor period is also observed. For a magnetic field
oriented along the $z$ axis (Fig.~2(f)), the polariton dynamics is relatively simple. Each hyperfine coherence $\hat{S}_{a~m}^{b~m}$ merely picks
up a phase factor that oscillates at frequency $2m\vert g_b\mathbf\Omega\vert$, thus returning the system to its initial state at half the Larmor
period.  In this case, the partial revival is actually a full revival.  On the other hand, for $\theta=\pi/2$ (Fig.~2(j)), a rotation through
half the Larmor period causes the coherence transformation $\hat{S}_{a~m}^{b~m} \rightarrow -\hat{S}_{a~-m}^{b~-m}$, and as a result, the
retrieval efficiency is reduced to $(\sum_m R_mR_{-m} / \sum_m R_m^2)^2$. The substructure within a half Larmor period is associated with
interference of different hyperfine coherences contributing to the dark-state polariton \cite{jenkins}.

\begin{figure}[htp]
\begin{center}
\leavevmode  \psfig{file=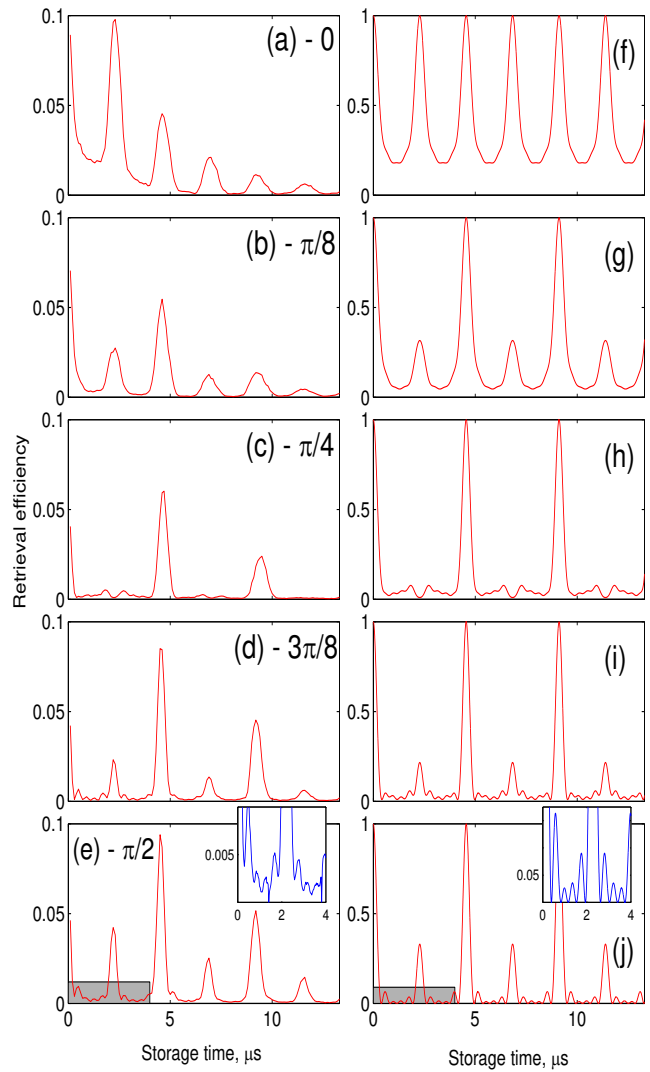,height=5.6in,width=3.4in}
\end{center}
\caption{ Panels (a)-(e) show the ratio of the number of photoelectric detection events for the retrieved and incident signal fields for various
orientations, $\theta =0,\pi/8, \pi/4, 3\pi/8, \pi/2$, of the applied magnetic field, and as a function of storage time. The incident signal
field is a weak coherent laser pulse. In all cases the control pulse is switched off at $T_s=0$. We observe a series of collapses and revivals at
multiples of the half Larmor period of $2.3$ $\mu$s. The observed damping over several Larmor periods is likely caused by residual magnetic field
gradients. The inset in Panel (e) demonstrates the observed substructure within the first Larmor period. Panels (f) through (j) are corresponding
theoretical plots of the dark-state polariton number calculated using Eq.(2).}\label{F2}
\end{figure}

To test these predictions, we apply a uniform dc magnetic field of magnitude $0.5 \pm 0.05$ G to the atomic ensemble using three pairs of
Helmholtz coils. In our first set of measurements, 150 ns long coherent laser pulses containing on average $\simeq 5$ photons serve as the signal
field. The outputs of the single-photon detectors D2 and D3 are fed into two ``Stop" inputs of a time-interval analyzer which records the arrival
times with a 2 ns time resolution. The electronic pulses from the detectors are gated for the period $[t_0,t_0 +T_g]$, with $T_g=240$ ns,
centered on the time determined by the control laser pulse during the retrieval stage. Counts recorded outside the gating period are therefore
removed from the analysis. The recorded data allows us to determine the number of photoelectric events $N_2+N_3$, where $N_i$ is the total number
of counts in the $i$-th channel($i=1,2,3$).

By measuring the retrieved field for different storage times and orientations of the magnetic field, we obtain the collapse and revival signals
shown in Fig.~2, (a) through (e). As expected, we observe  revivals at integer multiples of the Larmor period. In addition, we see partial
revivals at odd multiples of half the Larmor period, except in the vicinity of $\theta =\pi/4$. The measured substructures within a single
revival period are in good agreement with the theory (cf., insets of Fig.~2, (e) and (j)).  We attribute the overall damping of the revival
signal in the experimental data to the magnetic field gradients. The evident decrease of this damping while $\theta $ is varied from $0$ to
$\pi/2$ suggests that such gradients are predominantly along the direction of the signal field wavevector. Similarly, we attribute the additional
broadening of the revival peaks at longer times to inhomogeneous magnetic fields, possibly ac fields, not included in the theoretical
description. We are pursuing additional investigations to determine the temporal and spatial characteristics of the residual magnetic fields.

\begin{figure}[htp]
\begin{center}
\leavevmode  \psfig{file=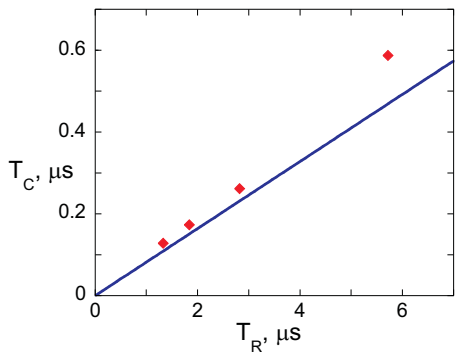,height=1.8in,width=2.34in}
\end{center}
\caption{Diamonds show the measured collapse time $T_C$ of the first revival at half the Larmor period as a function of the measured revival time
$T_R$, for magnetic field values of 0.8, 0.6, 0.4, and $0.2~\mbox{G}$, respectively, and for fixed orientation $\theta =\pi/2$. The line shows
the corresponding theoretical prediction $T_C\approx 0.082 T_R$  from Eq.(2).}\label{F2}
\end{figure}

Theory predicts that both the collapse and the revival times ($T_C$ and $T_R$, respectively) scale inversely with the magnetic field
\cite{jenkins}. In Fig.~3 the theoretical prediction $T_C\approx 0.082 T_R$ (solid line) is compared with the experimentally measured values. We
find very good agreement except for the lowest value of magnetic field  $B=0.2 ~\mbox{G}$ which may be explained by the presence of residual
magnetic field gradients.

\begin{figure}[htp]
\begin{center}
\leavevmode  \psfig{file=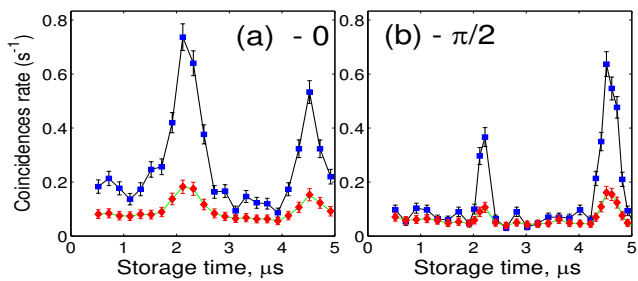,height=1.55in,width=3.4in}
\end{center}
\caption{Squares show the measured rate of coincidence detections between D1 and D2,3, $N_{si}=N_{12}+N_{13}$ as a function of the storage time.
Diamonds show the measured level of random coincidences $N_R$. The ratio of squares to diamonds gives $g_{si}$. Uncertainties are based on the
statistics of the photoelectric counting events.}\label{F3}
\end{figure}
In the measurements presented above, classical, coherent laser light was used as the signal field \cite{mandel}. We have also investigated the
revival dynamics of our atomic memory with the signal field in a single photon state, as shown in Fig.~1. The procedure for production of a
single photon state of the signal field at Site {\it A} is conditioned on the detection of an idler photon by D1 (see
Refs.\cite{matsukevich1,chaneliere} for further details). If photoelectric detections in different channels $1,2$ or $1,3$ happen within the same
gating period, they contribute to the corresponding coincidence counts between D1 and D{\it j}, $N_{1j}$, $j=2,3$. We evaluate the $\alpha $
parameter of Grangier {\it et al.} \cite{grangier}, given by the ratio of various photoelectric detection probabilities which are measured by the
set of detectors D1,D2 and D3. For an ideal single-photon state $\alpha =0$, whereas for coherent fields $\alpha =1$. We have experimentally
determined $\alpha =0.51 \pm 0.06$, without any correction for background or dark counts.

The normalized intensity cross-correlation function $g_{si} \equiv (N_{12}+N_{13})/N_{R}$ may be employed as a measure of non-classical
signal-idler field correlations \cite{mandel,walls}, as discussed in detail in Ref.\cite{kuzmich2}. Here $N_R\equiv N_1\cdot (N_2+N_3) \cdot
R_{rep}$ is the level of random coincidences, where $R_{rep}$ is the repetition rate of the experimental protocol. The values of $g_{si}$ are
obtained by the ratio of the upper and lower traces in Fig.~4. The measurements presented there give values of $g_{si}$ well in excess of two at
the revival times, suggesting the dark-state polaritons have a non-classical nature. One could further evaluate self-correlations for the idler
field $g_{ii}$, and for the signal field $g_{ss}$, and confirm that the Cauchy-Schwarz inequality $g_{si} ^{2} \leq g_{ss}g_{ii}$ is indeed
violated \cite{kuzmich2,mandel,walls}. We have measured, by adding a beamsplitter and an additional detector, the value $g_{ii}= 1.42 \pm 0.03$.
When the signal field is stored and retrieved 500 ns later, we find that both $g_{ss} \leq 2$ \cite{chaneliere}. While the total number of
recorded coincidences between detectors D2 and D3 is not high enough to evaluate $g_{ss}$ for the revived polariton, it is also expected to be
less than two, leading to a substantial violation of the Cauchy-Schwarz inequality.

In summary, we have demonstrated revivals of dark-state polaritons in a quantum memory element based on a cold atomic ensemble. The dynamical
manipulation and control of collective matter-field excitations, at the level of single quanta, is encouraging for further developments and
applications in quantum information science.

We thank M. S. Chapman for discussions and E.T. Neumann for experimental assistance. This work was supported by NASA, National Science
Foundation, Office of Naval Research, Research Corporation, Alfred P. Sloan Foundation, and Cullen-Peck Chair.

\end{document}